\documentclass[preprint,sort&compress,12pt]{elsarticle}

\pdfoutput=1

\usepackage{amssymb}
\usepackage{amsmath}
\usepackage{gensymb}
\usepackage{caption}
\usepackage{subcaption}
\usepackage{color}
\usepackage{xcolor}
\usepackage[normalem]{ulem}

\journal{Acta Materialia}

\begin{document}

\begin{frontmatter}

\title{Grain rotation and coupled grain boundary motion in two-dimensional binary hexagonal materials \tnoteref{t1}}
\tnotetext[t1]{\copyright~2021. This manuscript version is made available under the CC BY-NC-ND license
  https://creativecommons.org/licenses/by-nc-nd/4.0/}

\author[1]{Brendon Waters}
\author[1]{Zhi-Feng Huang\corref{cor1}}
\ead{huang@wayne.edu}
\cortext[cor1]{Corresponding author}
\address[1]{Department of Physics and Astronomy, Wayne State University, 
Detroit, Michigan 48201, USA}

\begin{abstract}

The dynamical mechanisms underlying the grain evolution and growth are of fundamental importance in controlling the structural properties of large-scale polycrystalline materials, but the effects of lattice ordering and distinct atomic species in multi-component material systems are still not well understood. We study these effects through the phase field crystal modeling of embedded curved grains in two-dimensional hexagonal materials, by examining and comparing the results of grain rotation, shrinking, and grain boundary dynamics over the full range of misorientation in binary systems of hexagonal boron nitride and single-component graphene monolayers. Calculations of the relation between grain radius and misorientation angle during time evolution reveal the normal-tangential coupled motion of the grain boundary matching the Cahn-Taylor formulation, as well as the transition to sliding and the regime of grain motion without rotation. The key effect of two-component sublattice ordering is identified, showing as a dual behavior of both positive and negative coupling modes with grains rotating towards increasing and decreasing angles, which is absent in two-dimensional single-component systems. The corresponding mechanisms are beyond the purely geometric considerations and require the energetic contribution from the difference between heteroelemental and homoelemental atomic bondings and the subsequent availability of a diverse variety of defect core structures and transformations. This indicates the important role played by the lattice inversion symmetry breaking in binary or multi-component materials, causing the change of detailed microstructures and dynamics of dislocation defects at grain boundaries as compared to single-component materials.

\end{abstract}


\end{frontmatter}

\section{Introduction}
\label{sec:intro}

In recent years there has been much excitement about the structures and properties of novel two-dimensional (2D) materials. With high mechanical strength, unique electronic properties, and tunable chemical reactivity \cite{Yazyev14,ChengNanoscale21} these materials offer promise in a wide variety of technological applications. These properties are a direct consequence of the 2D crystalline structure of the material, and thus to take full advantage of them requires the growth and synthesis of large-area, defect-free samples, a feat that has proved challenging. These large-scale samples are often grown by techniques of heteroepitaxy, such as chemical vapor deposition (CVD). While large areas can be covered with atomically thin layers of the desired material via these techniques, the resulting monolayer film is typically polycrystalline, as a result of spontaneous nucleation at various locations on the substrate and the formation and evolution of topological defects including dislocations and grain boundaries as observed in experiments \cite{Yazyev14,Huang11,GibbJACS13,RenNanoLett19,RenACSNano20,LinACSNano15,MendesACSNano19}.

It is thus desirable to remodel the material structure through post-grown processing such as annealing to facilitate the procedure of grain growth for achieving the single crystalline state of the system. Usually such a process begins with a large number of discreet grains separated by well-defined grain boundaries comprised of arrays of dislocation defects. The dynamics of these defects, particularly dislocation glide and climb, result in the motion of grain boundaries, with some grains growing at the expense of others. A large amount of research efforts, both experimentally \cite{Yazyev14,Huang11,GibbJACS13,RenNanoLett19,RenACSNano20,LinACSNano15,MendesACSNano19} and theoretically/computationally \cite{Taha17,HirvonenPRB16,LiJMPS18,Zhou19,MomeniNPJCM20}, have been put on the studies of defect dynamics and grain growth as well as the understanding and control of the corresponding mechanisms in various 2D material systems including graphene \cite{Yazyev14,Huang11,HirvonenPRB16,LiJMPS18,Zhou19}, hexagonal boron nitride (h-BN) \cite{GibbJACS13,Taha17,RenNanoLett19,RenACSNano20}, and transition metal dichalcogenides (TMDs) \cite{Yazyev14,LinACSNano15,MendesACSNano19}.

The mechanisms of grain growth, in both two- and three-dimensional (3D) materials, involve many coupled and competing factors in play \cite{CahnActaMater04,CahnActaMater06,CahnPhiloMag06,TrauttActaMater12a,TrauttActaMater12b,ThomasNatCommun17,WuActaMater12,McReynoldsActaMater16,AdlandPRL13}, resulting in complex phenomena of grain boundary migration and grain rotation. It has been shown that the migration of a grain boundary in its normal direction is geometrically coupled to the tangential motion of the grain along the boundary \cite{CahnActaMater04}, which tends to rotate the annealing grain. The strength and multiplicity of this coupling, as well as the degree of the accompanying shear deformation, depend on the misorientation of crystalline lattice across the grain boundary and the material growth or processing conditions, creating a rich landscape of possible motions as grains of different initial sizes and orientations interact. Properties of this type of shear-coupled motion have been studied via molecular dynamics (MD) \cite{CahnActaMater06,CahnPhiloMag06,TrauttActaMater12a,TrauttActaMater12b,ThomasNatCommun17} and phase field crystal (PFC) \cite{TrauttActaMater12b,WuActaMater12,McReynoldsActaMater16,AdlandPRL13} simulations, for both symmetric \cite{CahnActaMater06,CahnPhiloMag06,ThomasNatCommun17} and asymmetric \cite{TrauttActaMater12b} planar tilt grain boundaries in bicrystals, 3D cylindrical \cite{TrauttActaMater12a} or 2D circular \cite{WuActaMater12,McReynoldsActaMater16,AdlandPRL13} individual embedded grains, and multigrain systems of polycrystals \cite{McReynoldsActaMater16,ThomasNatCommun17}. The dual behavior or the switching between two different coupling modes has been found within a narrow transition range of misorientation angles in some 3D MD simulations of fcc metals like Cu \cite{CahnActaMater06,CahnPhiloMag06} and Ni \cite{ThomasNatCommun17} for the shearing of symmetric planar grain boundaries. It is noted that all these previous works are for single-component systems, while the modeling and understanding of the coupled grain boundary motion and grain rotation for binary or multi-component systems are still lacking.

In principle, this mechanism of normal-tangential coupled motion for grain growth dynamics is expected to directly apply to graphene-type 2D materials. However, not all 2D materials are created equal. Graphene is a widely studied conductive 2D carbon allotrope, with its six-fold symmetry responsible for its protected conducting states. A related material is 2D h-BN, which adopts the same honeycomb lattice-site structure as graphene, but with alternating B and N atoms occupying the neighboring lattice sites. This breaks the $60\degree$ symmetry of rotation for honeycomb lattice. Such a property of lattice polarity and inversion symmetry breaking occurs in all other 2D binary hexagonal materials such as TMDs (in the form of $MX_2$). If the shear-coupled mechanism of grain boundary motion would be purely geometric and dominated by the lattice structure and crystallographic characteristics, as assumed in previous studies \cite{CahnActaMater04,CahnActaMater06,CahnPhiloMag06,TrauttActaMater12a,TrauttActaMater12b,ThomasNatCommun17}, the property of the coupling between normal and tangential motions and the grain rotation induced would then be qualitatively similar for single-component graphene and two-component h-BN or TMDs (other than a simple extension of the full range of misorientation from $60\degree$ to $120\degree$), given that they have the same set of honeycomb lattice sites if neglecting the distinction between atomic species. However, this is not the case, as will be shown in this study.

In this paper we examine the dynamics of curved, embedded grains in both single-component and binary 2D hexagonal material systems, through the PFC modeling \cite{Elder04,Elder07,Taha17} which enables the study on diffusive timescales that are not accessible to atomistic simulations while still maintaining the microscopic resolution of individual defect structures. Our focus is on the effect of coupling between grain boundary normal motion and tangential translation in both systems, via the quantitative study of time variations of misorientation angle and grain radius during grain rotation, shrinking, and boundary evolution, and the matching to the analytic results of Cahn-Talyor formulation. Particularly important is the dual behavior, showing as the coexistence and switching between two different coupling modes, which is found in our simulations of h-BN type binary systems but absent in 2D single-component ones. The underlying atomic mechanisms are related to diverse types of dislocation core microstructures and their transformations that are made available in binary materials, indicating the pivotal role played by the inversion symmetry breaking and sublattice ordering in material systems composed of more than one species of constituents.

\section{Model and theory}

\subsection{The phase field crystal models}
\label{PFC}

We use a binary PFC model with sublattice ordering to study 2D monolayers of two-component hexagonal materials like h-BN, and for comparison, a single-component PFC model to simulate the graphene-type 2D system. The PFC models can be derived from classical density functional theory (DFT) by expanding the free energy functional $\mathcal{F}$ in terms of an atomic number density variation field $n(\vec{r},t)$ \cite{Elder07,re:teeffelen09,Huang10,Wang18R,Taha19}, allowing for spatially periodic solutions of the density field to represent the crystalline solid state of the material. The specific form of this free energy functional can be chosen to permit solutions with a desired crystalline lattice symmetry, incorporating system elasticity and plasticity, and be parameterized to match a real material of interest. The PFC method has the advantage of allowing the modeling of material systems with atomic-scale spatial resolution, while being computationally efficient enough to cover relatively long, diffusive timescales to examine complex dynamical phenomena such as defect migration and grain growth \cite{re:berry06,BerryPRB15,re:stefanovic09,BackofenActaMater14,TrauttActaMater12b,AdlandPRL13,WuActaMater12,McReynoldsActaMater16,Taha17,HirvonenPRB16,LiJMPS18,Zhou19,SkaugenPRB18,SkaugenPRL18,Salvalaglio20,Salvalaglio21}.

In the original, single-component PFC model, the dimensionless free energy functional (after rescaling) is given by \cite{Elder04,Elder07}
\begin{equation} 
\label{graphene energy}
\mathcal{F}=\int \mathrm{d} \vec{r} \left [ -\frac{\epsilon}{2} n^2 + \frac{1}{2} n (\nabla^2 + q_0^2)^2 n - \frac{g}{3} n^3 + \frac{1}{4} n^4 \right ],
\end{equation}
where $\epsilon$ and $g$ are phenomenological parameters that can be connected to the Fourier components of direct correlation functions in classical DFT \cite{Elder07,re:teeffelen09,Huang10}, with $\epsilon > 0$ required to produce a solid crystal with an appropriate choice of the average density variation $n_0$. The structure length unit has been rescaled such that the characteristic wave number $q_0 = 1$ for lattice periodicity. The dynamical evolution of the density field is governed by
\begin{equation}
\label{single PFC}
\frac{\partial n}{\partial t} = - \frac{\delta \mathcal{F}}{\delta n} + \mu
= - \left [ -\epsilon n + (\nabla^2 + q_0^2)^2 n - g n^2 + n^3 \right ] + \mu.
\end{equation}
Here the nonconserved dynamics with a constant chemical potential $\mu$ is used to model the evolution of grains during the growth and annealing process that resembles the experimental conditions. Experimentally, during the fabrication of 2D materials using epitaxial techniques (e.g., CVD), the constant flux is maintained under high-temperature growth conditions, where the sample is typically surrounded by gas-form surplus atoms subjected to specific chemical potential. As discussed in Sec.~\ref{sec:intro}, this growth process typically produces polycrystalline samples, grains of which then interact and evolve. The nonconserved dynamics, with the imposed constraint of pre-determined constant chemical potential for each atomic specie, is to model these conditions with constant flux. This type of nonconserved PFC dynamics (i.e., in the grand canonical ensemble) has also been used in some previous works simulating e.g., 2D single-component grain evolution and rotation \cite{AdlandPRL13}, shear-coupled motion of 2D grain boundaries \cite{TrauttActaMater12b}, and the evolution of $60\degree$ inversion domain boundaries in h-BN \cite{Taha17}, which all showed physically consistent results of dislocation motion as compared to those from conserved dynamics and also MD simulations (see e.g., Ref.~\cite{TrauttActaMater12b}).

Note that this single-component PFC system is invariant with respect to $n \rightarrow -n$ and $g \rightarrow -g$. It is straightforward to show that the corresponding solid phase in 2D is of triangular symmetry if $g-3n_0>0$, and when $g-3n_0<0$ it is of honeycomb lattice structure which is basically the inverse of triangular one, where the density maxima and minima have been reversed while the system elastic properties are maintained. This property has been utilized for the PFC modeling and the related model parameterization for graphene, including the structure, energy, and dynamics of dislocations, grain boundaries, polycrystals, and heterostructures \cite{HirvonenPRB16,LiJMPS18,Zhou19,HirvonenPRB19,ElderPRM21}. Although some elastic constants determined by this simple version of PFC model, particularly the value of Poisson ratio \cite{HirvonenPRB16}, are different from that of graphene, it would affect the properties of mechanical deformation of the sample but is not expected to qualitatively influence the behaviors of defect dynamics examined in this work. There would always be some quantitative differences which however should be small, given that any structural deformations/distortions caused by grain boundary and dislocation motion and interaction are local and of relatively small degree, and that the normal-tangential coupled motion of grain boundary, which is the main focus of this study, is related to local shear deformation of the lattice \cite{CahnActaMater04,CahnActaMater06,CahnPhiloMag06}.

In our simulations here the model parameters $\epsilon=0.02$, $g = -0.5$, $\mu=0.03$, and an initial value of $n_0=0.04$ are used. We chose the chemical potential value of our model by first numerically solving the corresponding PFC equation with conserved dynamics, with the use of the same model parameters in a perfect single crystal, allowing it to come to equilibrium, and then calculating the corresponding chemical potential $\mu$. We used this value of $\mu$ as the imposed constraint for our simulations with nonconserved dynamics, so that it would maintain the corresponding average density $n_0$ for a solid crystal. The high temperature regime our simulations operated in has a fairly narrow range of $n_0$ where the solid phase is stable, and thus this choice of chemical potential is important to maintain the solid state. The constraint of constant chemical potential then controls the global conservation of density in the system, given that the system simulated here is always in a solid phase with no liquid-solid interface involved. To verify this, we calculated the average density at time intervals throughout our simulations. After fluctuating during the initial transient time the average density of our PFC systems evolved to and was maintained at its own equilibrated value (with very small (less than $10^{-3}$) variations with time for this single-component PFC and also the binary system described below).

The 2D binary hexagonal materials can be effectively modeled by a two-component PFC model developed recently \cite{Taha17,Taha19}, where the sublattice ordering of $A$ and $B$ components is described by a rescaled free energy functional for
the density variation fields $n_A$ and $n_B$, i.e.,
\begin{equation} 
\label{hBN energy}
\begin{split}
\mathcal{F} = \int \mathrm{d} \vec{r} & \left [ -\frac{1}{2} \epsilon_A n_A^2 + \frac{1}{2} n_A(\nabla^2 + q_A^2)^2n_A - \frac{1}{3} g_A n_A^3 + \frac{1}{4} n_A^4 \right. \\
& - \frac{1}{2} \epsilon_B n_B^2 + \frac{1}{2}\beta_B n_B(\nabla^2 + q_B^2)^2 n_B - \frac{1}{3} g_B n_B^3 + \frac{1}{4} v n_B^4 \\
& \left. + \alpha_{AB} n_A n_B + \beta_{AB} n_A (\nabla^2 + q_{AB}^2)^2 n_B + \frac{1}{2} w n_A^2 n_B +\frac{1}{2} u n_A n_B^2 \right ].
\end{split}
\end{equation}
The corresponding PFC dynamical equations are given by
\begin{eqnarray}
\frac{\partial n_A}{\partial t} &=& - \frac{\delta \mathcal{F}}{\delta n_A} + \mu_A = - \left [ -\epsilon_A n_A + \left (\nabla^2 + q_A^2 \right )^2 n_A - g_A n_A^2 + n_A^3 \right. \nonumber\\
&& \left. + \alpha_{AB} n_B + \beta_{AB} \left (\nabla^2 + q_{AB}^2 \right )^2 n_B + w n_A n_B + \frac{u}{2} n_B^2 \right ] + \mu_A, \label{binary PFC 1}\\
\frac{\partial n_B}{\partial t} &=& -m_B \left ( \frac{\delta \mathcal{F}}{\delta n_B} - \mu_B \right ) = -m_B \left [ -\epsilon_B n_B + \beta_B \left (\nabla^2 + q_B^2 \right )^2 n_B - g_B n_B^2 \right. \nonumber\\
&& \left. + vn_B^3 + \alpha_{AB} n_A + \beta_{AB} \left (\nabla^2 + q_{AB}^2 \right )^2 n_A + u n_A n_B + \frac{w}{2}n_A^2 - \mu_B \right ]
\label{binary PFC 2}
\end{eqnarray}
where $m_B$ is the mobility ratio between $B$ and $A$ components, $\mu_A$ and $\mu_B$ are chemical potential values of $A$ and $B$ species, and the dimensionless model parameters $\epsilon_{A(B)}$, $q_{A(B)}$, $q_{AB}$, $g_{A(B)}$, $\alpha_{AB}$, $\beta_{AB}$, $\beta_B$, $v$, $w$, and $u$ can be expressed in terms of the Fourier-expansion coefficients of two- and three-point direct correlation functions in classical DFT for a binary $AB$ system \cite{Taha19}. In Eqs.~(\ref{hBN energy}) and (\ref{binary PFC 1}), the first four terms correspond to the single-component description for component $A$ with density field $n_A$, while the next four terms in Eq.~(\ref{hBN energy}) and the first four terms in Eq.~(\ref{binary PFC 2}) are for component $B$. All the other terms represent the coupling between $A$ and $B$ species, ensuring no overlap between $A$ and $B$ density maxima and the stabilization of a binary honeycomb structure, which consists of two $A$ and $B$ triangular sublattices breaking the inversion symmetry of the overall binary lattice (i.e., with 3 triangular lattice sites occupied by component $A$ and the other 3 occupied by $B$ in a 6-membered honeycomb unit ring). Also importantly, the model free energy functional is constructed to energetically favor the $A$-$B$ heteroelemental neighboring with respect to the $A$-$A$ or $B$-$B$ homoelemental ones, as occurred in real two-component materials.

In binary PFC equations (\ref{binary PFC 1}) and (\ref{binary PFC 2}) the nonconserved dynamics with constant chemical potentials is again used to simulate the grain growth with constant flux. In addition to better modeling the growth conditions, an advantage of using this nonconserved dynamics is related to the mobility of embedded grains in the binary PFC. We found that when applying the more conventional conserved dynamics (with the absence of thermal noise in our study), the embedded grains of binary PFCs have rather low mobility due to large Peierls barrier for defect motion and rotate too little to measure, while the nonconserved dynamics appeared to ease this problem. We thus used the nonconserved dynamics for both binary and single-component PFCs so that they can be directly compared with as few confounding factors as possible. We also chose initial conditions that resulted in high enough mobilities of defects along the grain boundary and therefore large enough degrees of grain rotations for analysis.

This binary PFC model has been parameterized for 2D h-BN monolayers and been applied to recent studies of h-BN grain boundary structures and dynamics \cite{Taha17}, thermal transport in pristine and polycrystalline h-BN layers \cite{DongPCCP18}, and both graphene/h-BN and h-BN/h-BN vertical heterostructures \cite{HirvonenPRB19,ElderPRM21}. In the following calculations we use parameters $\epsilon_A=\epsilon_B=0.02$, $q_A=q_B=q_{AB}=1$, $v=\beta_B=1$, $\alpha_{AB}=0.5$, $\beta_{AB}=0.02$, $g_A=g_B=0.5$, $w=u=0.3$, $m_B=1$, $\mu_A=\mu_B=-0.45$, and the average density variations $n_{A0}=n_{B0}=-0.4$. By matching to the energy and length scales of 2D h-BN, it has been shown that a unit length in this model corresponds to $0.342$ {\AA} and an energy unit is of 2.74 eV \cite{Taha17}. In principle, out-of-plane deformations can be incorporated in the model (as implemented very recently \cite{ElderPRM21}), the effect of which is neglected in this study as here we focus on the evolution of monolayer grains confined on a substrate (e.g., during CVD growth), where very small or flattening vertical variation of epitaxial overlayers was observed in experiments of h-BN \cite{GibbJACS13} and thus plays a negligible or secondary role.

\subsection{The Cahn-Taylor formulation for grain boundary coupled motion}

Qualitatively, an embedded grain within a background crystal or matrix is a section of that crystal that has its lattice planes misaligned by some angle $\theta$ relative to the orientation of the background lattice. This misalignment causes the formation of an enclosed, curved grain boundary surrounding the grain, a region where the crystalline lattice changes orientation, punctuated by lattice dislocation defects which accommodate this angle change. Despite this misalignment, lattice planes of the embedded grain tend to maintain continuity with those of the surrounding matrix across the grain boundary. This produces an effective coupling between the normal motion of the grain boundary and tangential translation along the boundary, which, repeated all along the circumference of the grain, would then induce a net rotation of the grain during grain shrinkage or growth to maintain the lattice-plane continuity \cite{CahnActaMater04}. This process of coupled motion has been found to occur not only for small-angle grain boundaries with arrays of discrete boundary dislocations, but also for high-angle boundaries with connected defects, with the direction of boundary motion or grain rotation (i.e., towards larger or smaller misorientation angle $\theta$) depending on the initial misorientation and the coupling mode \cite{CahnActaMater06,CahnPhiloMag06,TrauttActaMater12a,TrauttActaMater12b,ThomasNatCommun17}. 

As in the formulation of Cahn and Taylor \cite{CahnActaMater04}, if the grain changes size due to motion of the grain boundary normal to the grain-matrix interface at velocity $v_n$, it will tend to cause a tangential motion of the grain at velocity $v_\parallel$ to accommodate the continuity of the lattice planes. In addition, tangential motion could be induced by a shear stress $\sigma$ without the coupling to normal motion, referred to as sliding. Thus, the combined tangential motion of the grain gives $v_\parallel = \beta v_n + S \sigma$, where $\beta$ is the coupling factor and $S$ is the sliding coefficient. In general, these coefficients can be functions of $\theta$. For a cylindrical or circular embedded grain with radius $r$, the general equations of motion have been derived by Cahn and Taylor \cite{CahnActaMater04} as
\begin{equation} 
\label{vperp}
    v_n = -\frac{dr}{dt} = M \left( \frac{\gamma - \beta \gamma'}{r} + \beta \sigma \right),
\end{equation}
\begin{equation} 
\label{vparallel}
    -v_\parallel = r \frac{d \theta}{dt} = \beta M \left( \frac{\gamma - \beta \gamma'}{r} + \beta \sigma \right) - S \left(\frac{\gamma'}{r} - \sigma \right),
\end{equation}
and from dividing Eq.~(\ref{vparallel}) by Eq.~(\ref{vperp}),
\begin{equation} 
\label{full_dynamics}
    \frac{d \theta}{d \ln{r}} = - \beta + \frac{S(\gamma'/r - \sigma)}{M [(\gamma - \beta \gamma')/r + \beta \sigma]},
\end{equation}
where $M$ is the mobility of grain boundary migration, $\gamma$ is the grain boundary energy, $\gamma'=d\gamma/d\theta$, and the absence of grain-matrix bulk free energy difference has been assumed. Thus, Eq.~(\ref{full_dynamics}) indicates that the grain motion can be described by a relationship between $\theta$ and $\ln r$ controlled by the strength of coupling (i.e., $\beta$) and sliding (i.e., $S$).

It has been hypothesized that the coupling occurs or dominates in most cases of grain boundary motion, as verified in recent MD and PFC simulations \cite{CahnActaMater06,CahnPhiloMag06,TrauttActaMater12a,TrauttActaMater12b,ThomasNatCommun17,WuActaMater12,McReynoldsActaMater16,AdlandPRL13}, except in some narrow ranges of $\theta$ corresponding to symmetry points of the underlying crystalline lattice where $\beta \rightarrow 0$ \cite{CahnActaMater04}. While the contribution of sliding may be significant for some ranges of $\theta$, it would be difficult to determine quantitatively through coefficient $S$ which is generally unknown. It is therefore more practical to measure sliding as a deviation from the idealized coupling behavior \cite{CahnActaMater06,TrauttActaMater12a,TrauttActaMater12b}. For example, for pure or perfect coupling with $S = 0$, Eq.~(\ref{full_dynamics}) reduces to a geometric relation
\begin{equation} 
\label{coupling}
    \frac{d \theta}{d \ln{r}} = - \beta.
\end{equation}
Once the $\theta$ dependence of $\beta$ is known, integrating Eq.~(\ref{coupling}) would yield a master curve \cite{CahnActaMater04} governing the relation between grain radius $r$ and misorientation $\theta$. The deviation of the behavior of an embedded grain from this ideal coupling relation can then be used to identify the degree of sliding.

\subsection{Multiplicity of the coupling factor in 2D hexagonal systems}
\label{sec:multimodes}

Recent studies based on both geometric dislocation or disconnection models and MD simulations \cite{CahnActaMater06,CahnPhiloMag06,ThomasNatCommun17} have shown that the coupling factor $\beta$ is a multivalued function of the misorientation angle $\theta$, with different branches/modes of $\beta(\theta)$. Because $\beta$ is a purely geometric coupling factor, it can be derived from geometric considerations of the grain boundary, giving $\beta(\theta) \propto \tan(\theta/2 - k\pi/N)$ with $k=0,1,2,...,N-1$ for $N$-fold rotational symmetry of the lattice \cite{CahnActaMater06,TrauttActaMater12b}. In the previous studies of coupled grain boundary motion, either 3D cubic or 2D square symmetry is considered, corresponding to $N=4$; only the first two modes, $k=0,1$, were observed, i.e., $\beta_1=\beta_{<100>}=2\tan(\theta/2)$ and $\beta_2=\beta_{<110>}=2\tan(\theta/2-\pi/4)$ \cite{CahnActaMater06,CahnPhiloMag06}, but not the other two which would involve too large angles.

For the 2D hexagonal systems examined here, we have six-fold symmetry (i.e., $N=6$). Following the similar procedure in Ref.~\cite{CahnActaMater06} based on the dislocation model, the coupling factor for the single-component honeycomb structure is obtained as
\begin{equation} 
\label{beta1}
    \beta_1 = 2 \tan \left( \frac{\theta}{2} \right).
\end{equation}
Substituting it into Eq.~(\ref{coupling}) and integrating over $\theta$ lead to (with an integration constant $C_1$)
\begin{equation} 
\label{lnr1}
    \ln r = - \ln{\left[\sin{\left(\frac{\theta}{2}\right)}\right]} + C_1,
\end{equation}
which can be used to obtain a master curve relating $r$ and $\theta$ for the embedded grain in the case of perfect coupling. 

The coupling factor $\beta_1$ in Eq.~(\ref{beta1}) represents a positive branch of coupling. Another branch, with negative coupling factor $\beta$ related to opposite direction of grain rotation, can be obtained by considering the $60\degree$ rotational invariance of the lattice structure, i.e., via $\theta \rightarrow \theta-\pi/3$ in Eq.~(\ref{beta1}), giving
\begin{equation} 
\label{beta2}
\beta_2 = - 2 \tan \left ( \frac{\pi}{6} - \frac{\theta}{2} \right ),
\end{equation}
which results in
\begin{equation} 
\label{lnr2}
    \ln r = - \ln{\left[\sin{\left(\frac{\pi}{6}-\frac{\theta}{2}\right)}\right]} + C_2,
\end{equation}
where $C_2$ is the integration constant. As stated above, in general there are $N=6$ possible branches of the multivalued function $\beta(\theta)$ due to six-fold symmetry of the honeycomb lattice, i.e., $\beta = 2 \tan{(\theta/2 - k \pi/6)}$ with $k = {0,1,2,3,4,5}$. Typically only the first two with smallest magnitudes, i.e., $k = 0,1$ corresponding to Eqs.~(\ref{beta1}) and (\ref{beta2}) respectively, are expected to be found in practice, while the others are too large to couple effectively to grains that can be physically realized. This has been seen in our PFC simulations of 2D single-component systems of graphene, as will be shown in the next section.

\begin{figure}[htb]
    \centering
    \includegraphics[width=0.8\textwidth]{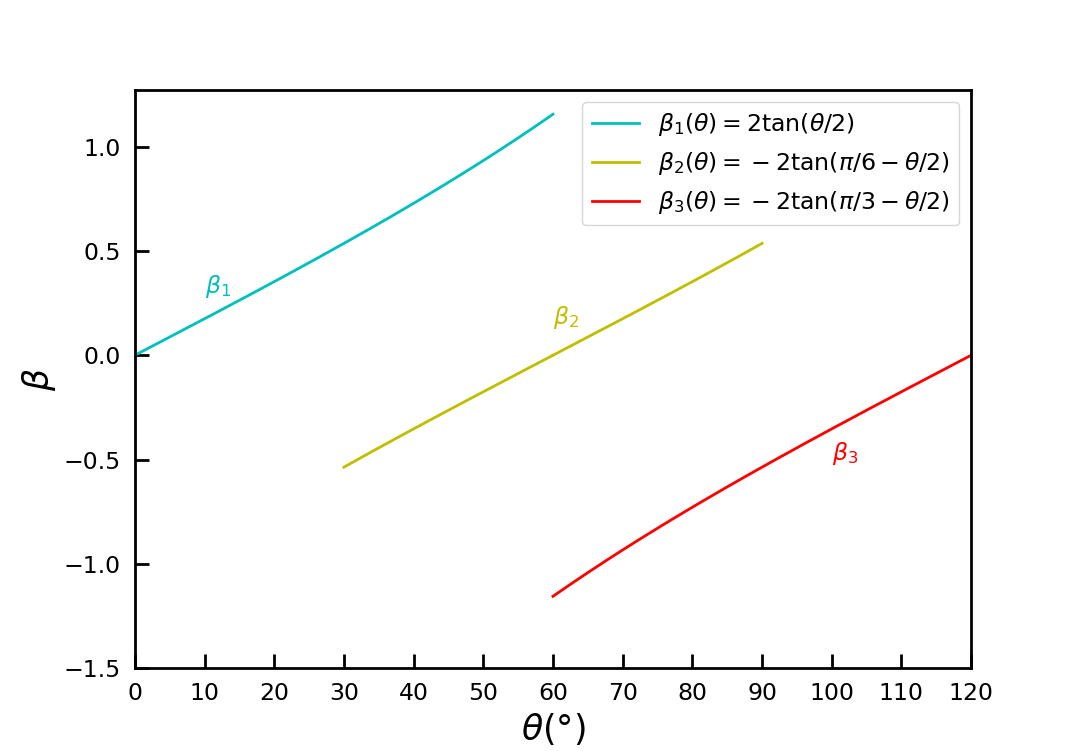}
    \caption{Branches of $\beta(\theta)$ for 2D binary hexagonal systems. Note that near the branching points of $\theta = 30 \degree$ and $\theta = 90 \degree$  the positive and negative branches of $\beta(\theta)$ approach the same magnitude.}
    \label{fig:beta_branches}
\end{figure}

For two-component 2D hexagonal materials like h-BN, the full range of grain boundary misorientation angles is from $0\degree$ to $120\degree$, instead of $60\degree$ for single-component systems like graphene, as a result of the binary $AB$ sublattice ordering and the subsequent lattice inversion symmetry breaking. For $\theta$ in the range of $[60\degree, 90\degree]$, we have $\beta = 2 \tan{(\theta/2 - \pi/6)}$, of the same form as Eq.~(\ref{beta2}) but now positive. The corresponding $r$ vs $\theta$ master curve then becomes
\begin{equation} 
\label{lnr3}
    \ln r = - \ln{\left[\sin{\left(\frac{\theta}{2}-\frac{\pi}{6}\right)}\right]} + C_3.
\end{equation}
The negative branch of $\beta$ for the $\theta$ range of $[60\degree, 120\degree]$ can be obtained by replacing $\theta \rightarrow \theta-2\pi/3$ in Eq.~(\ref{beta1}) or equivalently, $\theta \rightarrow \theta-\pi/3$ in Eq.~(\ref{beta2}), i.e.,
\begin{equation} 
\label{beta3}
\beta_3 = - 2 \tan \left ( \frac{\pi}{3} - \frac{\theta}{2} \right ),
\end{equation}
corresponding to $k=2$ mode. From Eq.~(\ref{coupling}) we have
\begin{equation} 
\label{lnr4}
    \ln r = - \ln{\left[\sin{\left(\frac{\pi}{3}-\frac{\theta}{2}\right)}\right]} + C_4.
\end{equation}
In Eqs.~(\ref{lnr3}) and (\ref{lnr4}) the integration constants $C_3$ and $C_4$ are determined by initial conditions. Thus, for a binary hexagonal system three coupling modes ($k=0,1,2$) can be identified, as shown in Eqs. (\ref{beta1}), (\ref{beta2}), (\ref{beta3}) and Fig.~\ref{fig:beta_branches}, resulting in four $r$ vs $\theta$ master curves determined by Eqs.~(\ref{lnr1}), (\ref{lnr2}), (\ref{lnr3}), and (\ref{lnr4}).

If the coupling mechanism between normal motion and tangential translation is attributed only to the geometric factors of lattice structure, as in the Cahn-Taylor formulation, the property of grain dynamics in the $\theta$ range of $[60\degree, 120\degree]$ is expected to mirror that of $[0\degree, 60\degree]$, resembling the behavior of single-component systems (e.g., graphene) which have the same honeycomb lattice. However, our simulation results indicate that this scenario based on purely geometric consideration is only partially true for a binary ordering system like h-BN, and additional factors due to the energetic contribution of binary components with inversion symmetry breaking and the resulting different defect structures need to be incorporated, leading to more complex behavior of grain motion as will be detailed below.

\section{Simulations and Results}

\subsection{Simulation method and setup}
\label{sec:setup}

We numerically solved the aforementioned PFC equations using a pseudospectral method in a square simulation box under periodic boundary conditions, with a resolution of 8 numerical grid points per lattice spacing. Although the numerical grid spacings $(\Delta x, \Delta y)$ can be chosen to minimize the system free energy of a single crystal with hexagonal structure, which can effectively remove most of strain in the bulk and fit integer number of atoms (density peaks) into the simulation box with equilibrium lattice spacing \cite{Taha17}, they would be of different values for single-component and binary systems simulated. For a more direct comparison between these two types of system it is preferable to set up their simulation conditions as close as possible, and thus we used the same values of $\Delta x=\Delta y=2\pi/8$, which inevitably introduced weak strain into the simulation box. The stress induced in the whole system by the simulation setup should make some small quantitative influence on the grain motion but would not affect the overall results of grain boundary coupled motion examined in this work, given that grains of all the angles studied were set up with the same conditions. In addition, in polycrystalline samples of real materials each grain is usually subjected to the impact of stress field generated by the other grains due to long-range elastic interaction, and our simulation of single embedded grains under background strain could serve as a related model system.

The systems were initialized with the majority of the area in a perfect honeycomb lattice with the PFC density field $n$ or $n_A$ and $n_B$ approximated by the corresponding analytic expression in the one-mode approximation, to be further relaxed via the full PFC equations. This single crystal then had a circular area rotated by an initial angle $\theta_0$ to form an embedded grain, by projecting the $x$ and $y$ coordinates onto a new coordinate system rotated by $\theta_0$ and recalculating the approximated $n$ values for initialization. These embedded grains were allowed to evolve until they either ceased further motion, or shrank completely and disappeared with all the boundary defects annihilated.

Delaunay triangulation was used to identify the connection from each simulated atomic site to its nearest neighbors, for measuring the locations of defects and local lattice orientations. Since the inverse of honeycomb structure is triangular which is more convenient for analysis, here we use local density minimum (i.e., local minimum of $n$ or $n_A+n_B$, corresponding to the real vacancy site) as the ``atomic site'' in the triangulation to generate the Delaunay graph. If the site has exactly 6 neighbors, the angles of the triangulation edges were computed to calculate the local angle of the lattice in the neighborhood of that site. If the number of neighboring differs from 6, it was labeled as a disclination defect. At a given time step the overall misorientation angle $\theta$ of the embedded grain was determined by averaging over the local angles of a group of sites within a small radius corresponding to approximately three lattice periods near the center of the grain, so that the angle calculation would not be affected by the shrinking grain until nearly the final moment of grain collapse. As the grains shrank and evolved we occasionally observed narrow regions of local angle change in the background crystal outside the grain. Some of these distortions would be the result of strain induced by the simulation box setup, but they were small in terms of both magnitude and spatial extent, showing typically as narrow bands of 1-3 lattice periods wide which are faintly visible outside the grain (see e.g., Fig.~\ref{fig:binary_rotation}). In our setup the embedded grain was located far enough from the boundaries, and any defects which would nucleate near the edge of the strained simulation box were several lattice periods away from the grain and did not interact with the grain boundary in any obvious way observed.

The grains frequently deviated from their initial circular shape during the course of evolution, becoming irregular or faceted. Thus, to calculate an effective grain radius $r$ we constructed a polygon connecting the dislocations along the grain boundary and counted the number of sites within it, with the radius of the grain proportional to the square root of this number. This effectively averaged the radius around the circumference of the approximately circular grain.

To study the $\theta$ dependent property of embedded grain rotation, a large number of simulations were initialized at different initial angles $\theta_0$ and evolved with time. Each grain was initialized with a radius of 64 grid points, corresponding to a diameter of approximately 16 lattice spacings. This setup of relatively small initial grains is to better facilitate elastic relaxation and grain rotation, particularly for the binary PFC grains which exhibit much lower mobility and are more reluctant to rotate (due to higher Peierls barrier) as compared to the single-component ones. This limited grain size results in high curvature of the grain boundary and irregular grain shape during time evolution. For very low misorientation angles there would be few sparsely distributed dislocations around the irregular circumference of the curved grain. When the spacing of the defects along the grain boundary becomes too large as compared to the radius of curvature of the boundary, the boundary itself and the grain size become ill-defined and ambiguous. Grains in this state with widely spaced defects, which tended to act more like isolated dislocations rather than those of a distinct grain boundary, also tend to evolve slowly or even cease to evolve after initial transients. Furthermore, at these lowest misorientation angles the normal-tangential coupling is the weakest according to the Cahn-Taylor formulation (see Fig.~\ref{fig:beta_branches}). Therefore, in this study we set the minimum initial misorientation at $\theta_0 \approx 5 \degree$, and mainly focus on the behavior of embedded grains at larger misorientation angles.

To study different initial conditions at the same $\theta_0$ we shifted the initial grain center by small displacements along the horizontal and vertical directions, to create grain boundaries that intersect different sets of lattice planes. These displacements were set by integer numbers of numerical grid points, less than 8 points that cover one lattice period as the properties of the crystal would be periodic over this range.

We have examined the outcomes of simulated grain evolution to ensure the nonconserved PFC dynamics used in this work (with the imposed constraint of constant chemical potential) generated correct behaviors of defect dynamics. The number of dislocation defects remained conserved before their annihilation or recombination, which is consistent with that found in previous PFC study of grain rotation in 2D square lattice using this type of nonconserved dynamics \cite{AdlandPRL13}. Also, the process of defect annihilation and recombination was observed to occur only between neighboring dislocation rings but not at larger or arbitrary separation. A known issue of PFC simulations, which exists for both conserved and nonconserved dynamics, is that in some cases the number of atomic sites (density peaks) near a dislocation core would vary and thus be not strictly conserved during dislocation motion (e.g., climb), which can be attributed to the result of vacancy diffusion (see e.g., discussions in Refs.~\cite{McReynoldsActaMater16} and \cite{TrauttActaMater12b}). Additionally, we compared our simulation results to those of Ref.~\cite{WuActaMater12} using conserved dynamics for the study of single-component PFC with triangular lattice symmetry. Similar behavior of single-component embedded grain motion has been found using both types of dynamics, with no significant differences between them (see below for more details), indicating that conserved and nonconserved PFC dynamics give similar and consistent results of grain rotation.

\subsection{Embedded grains in single-component PFCs for graphene}

\begin{figure}[htbp]
    \centering
    \includegraphics[width=0.95\textwidth]{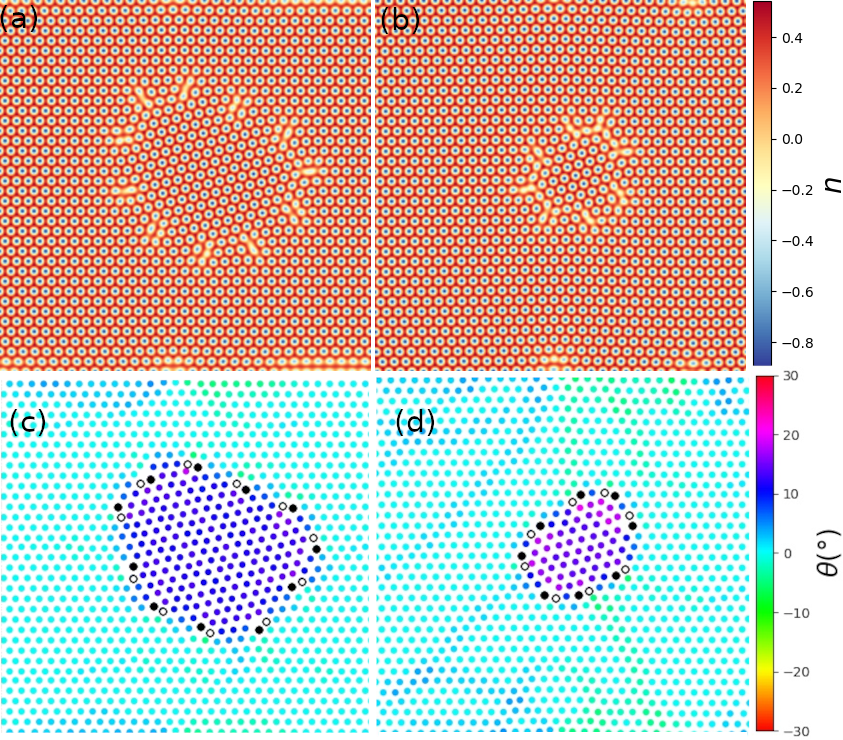}
    \caption{Snapshots of grain shrinking and rotation obtained from single-component PFC simulation, at early [(a) and (c)] and late [(b) and (d)] times of evolution. The grain was initialized with $\theta_0 = 12 \degree$. The spatial configurations of the density profile $n$ are shown in (a) and (b), with honeycomb lattice structure. The corresponding spatial distributions of local orientation angle $\theta$ are given in (c) and (d) as obtained from Delaunay triangulation, where each site represents the local minimum of $n$ colored by the value of local angle $\theta$. Sites with more than 6 neighbors are marked in black, and those with fewer than 6 neighbors marked in white, indicating the locations of defect cores. Note that the increase in the degree of purple coloring inside the grain from (c) to (d) represents the grain's rotation towards larger $\theta$, and the faint streamers of color outside the grain represent areas of angle change caused by strain in the crystal.}
    \label{fig:single_rotation}
\end{figure}

Single-component PFCs exhibit a honeycomb lattice structure with six-fold rotational symmetry, so that $\theta < \theta_{\rm max}^{\rm s} = 60 \degree$. Embedded grains in our PFC simulations of graphene systems based on Eq.~(\ref{single PFC}) tended to have relatively high mobility, often rotating across $5 \degree$ or more before grain collapsing. At low angles $(\theta < 15 \degree)$ they behaved as predicted by the Cahn-Talyor formulation for the coupling of normal-tangential motions, tending to decrease in radius while $\theta$ increased simultaneously, with some sample snapshots shown in Fig.~\ref{fig:single_rotation}. These embedded grains often underwent a faceting-defaceting transition, as described in Ref.~\cite{WuActaMater12} which examined the dynamics of circular grains in 2D PFC with triangular structure. At higher misalignment angles $(15 \degree < \theta < 30 \degree)$ grains tended to rotate much less or not at all, agreeing again with Ref.~\cite{WuActaMater12} and indicating that they experienced a substantial decrease in coupling. Details of the simulations (with the help of Delaunay triangulation for defects identification) showed pairs of dislocation cores around the boundary of the embedded grain would approach each other and finally annihilate when the spacing between them became sufficiently small. This would then limit the maximum density of dislocations, preventing any further increase in $\theta$. Grains initialized at or near $\theta = 30 \degree = \theta_{\rm max}^{\rm s}/2$ did not rotate, and decreased in radius without any obvious changes in the overall structure or orientation.

\begin{figure}[htbp]
    \centering
    \includegraphics[width=\textwidth]{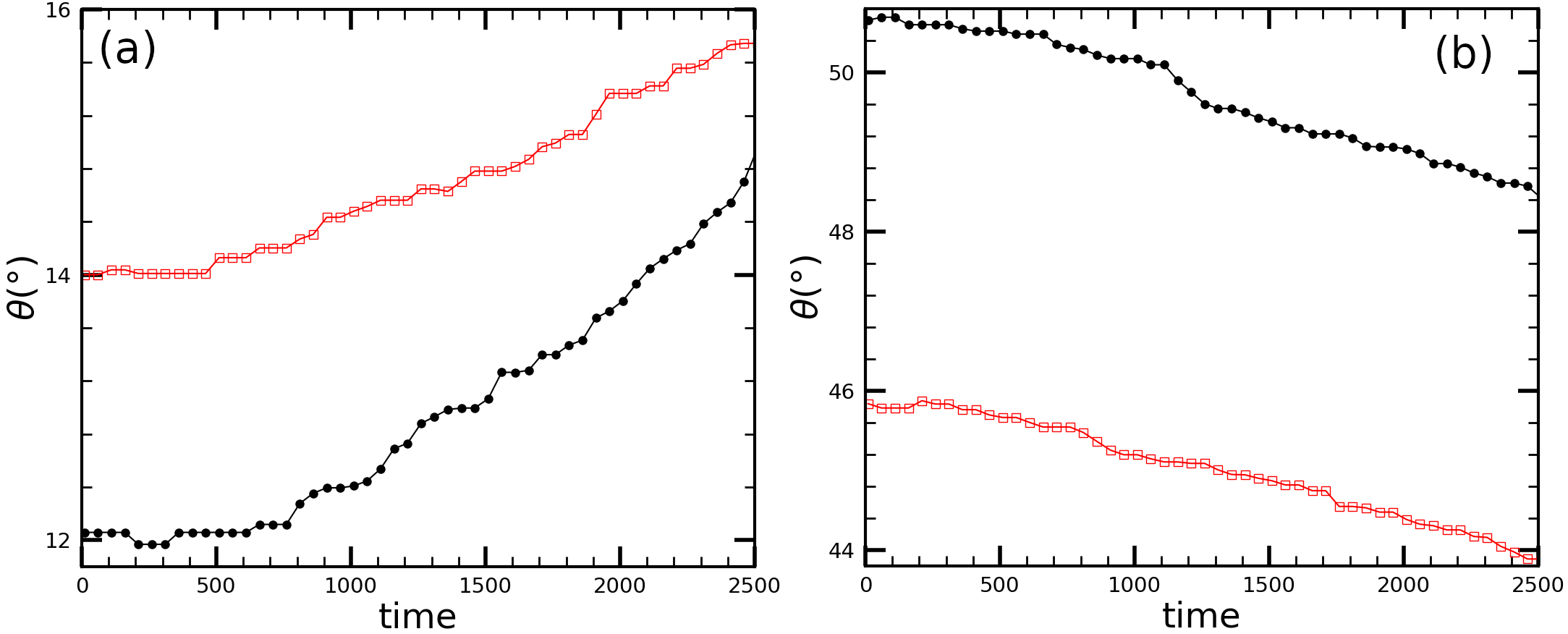}
    \caption{Grain misorientation angle $\theta$ as a function of time, for single-component embedded grains initialized at two different angles in each panel. (a) Grains were initialized in the low-angle regime around $12 \degree$ and $14 \degree$, and rotated towards increasing $\theta$ during time evolution and shrinking. (b) Grains were initialized in the high-angle regime around $46 \degree$ and $50 \degree$, and rotated towards lower $\theta$ instead.}
    \label{fig:single_angle_time}
\end{figure}

Results for $\theta > 30\degree$ are similar, mirroring those below $30\degree$ as expected from $60\degree$ rotational invariance of the single-component honeycomb lattice. The small misorientations now correspond to angles close to $60\degree$ (i.e., $\theta \rightarrow 60\degree - \theta$), with the grain rotating towards lower angle instead during its shrinkage. Some sample simulation results for the time evolution of the misorientation angle $\theta$ are given in Fig.~\ref{fig:single_angle_time}, showing both cases of grain rotation towards increasing or decreasing $\theta$ when starting from initial angles below or above $30\degree$, respectively. This is consistent with previous MD simulations giving similarly two types of time-dependent behavior of the misorientation angle for 3D cylindrical grain rotation in a fcc metal of Cu \cite{TrauttActaMater12a}. Note that some small nonmonotonic fluctuations occur at the early time range in Fig.~\ref{fig:single_angle_time}, due to initial transients of grain relaxation, while at later times the monotonic behavior of $\theta$, either increasing (Fig.~\ref{fig:single_angle_time}(a)) or decreasing (Fig.~\ref{fig:single_angle_time}(b)) with time, maintains, with few slight variations within the measurement uncertainty.

\begin{figure}[htbp]
    \centering
    \includegraphics[width=0.9\textwidth]{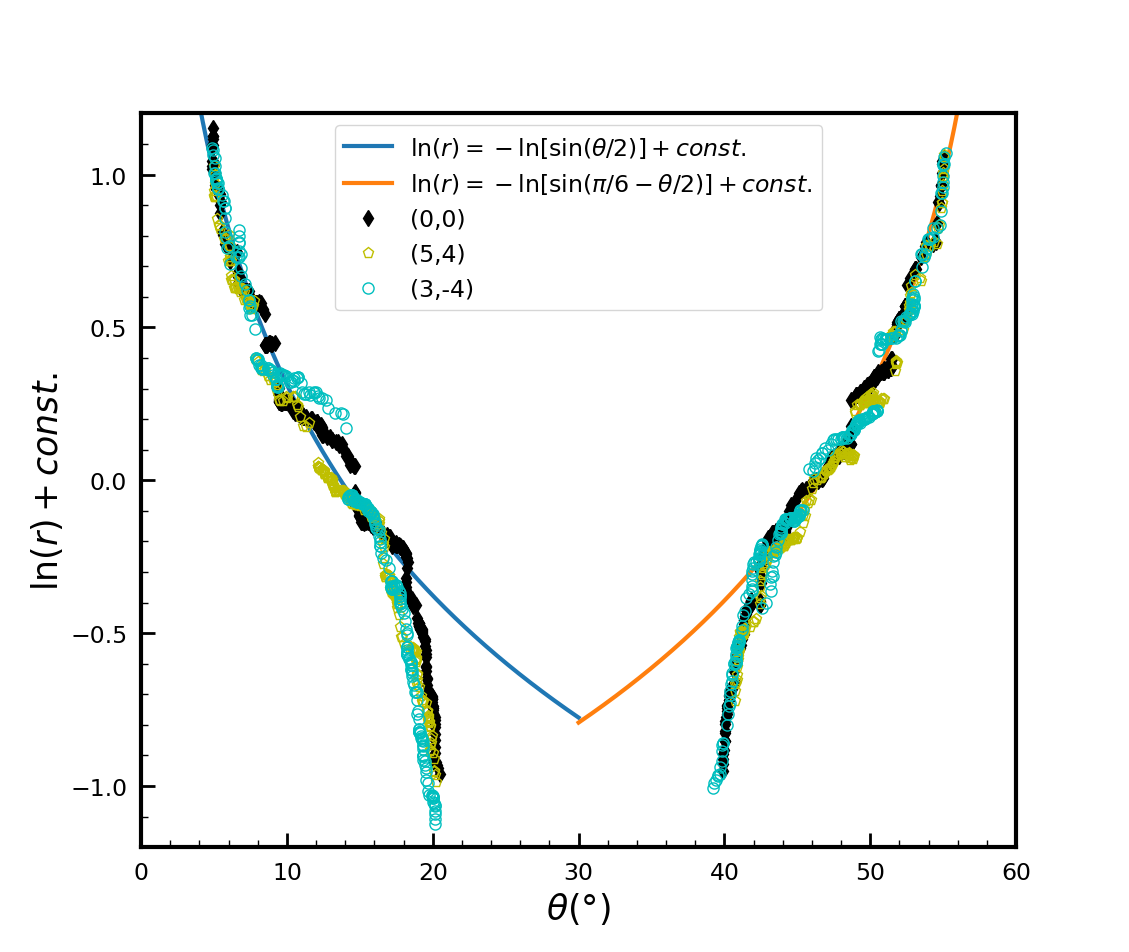}
    \caption{Plots of $\ln r$ vs $\theta$, including both the numerical data obtained from PFC simulations of single-component embedded grains and the analytic master curves (Eqs.~(\ref{lnr1}) and (\ref{lnr2})) for the two primary coupling branches of $\beta$. Different types of symbols represent the simulation results for grains initialized with different displacements of grain center relative to the background crystal, with the horizontal and vertical components of each displacement (in units of grid points) indicated.}
    \label{fig:single_rotation_lnr}
\end{figure}

These two types of grain rotation to opposite directions correspond to the two coupling modes described in Sec.~\ref{sec:multimodes}, with two branches of the coupling factor $\beta$ of opposite signs given in Eqs.~(\ref{beta1}) and (\ref{beta2}). To develop a more quantitative understanding of our simulation results, we measured the effective grain radius $r$ as a function of the evolving grain misorientation $\theta$ and plot the result of $\ln{(r(\theta))}$ in Fig.~\ref{fig:single_rotation_lnr}. Also plotted are the corresponding analytic expressions of Eqs.~(\ref{lnr1}) and (\ref{lnr2}), serving as the master curves \cite{CahnActaMater04} in the condition of perfect normal-tangential coupling. We conducted a series of PFC simulations by initializing the subsequent embedded grain with an initial angle $\theta_0$ equal to the final-stage angle of the previous grain simulation, and obtained sections of the $\ln r$ curve across all possible $\theta$. To match or fit the measured data with the master curves, we need to consider the unknown integration constants $C_1$ and $C_2$ in the analytic results of Eqs.~(\ref{lnr1}) and (\ref{lnr2}) by following the similar step of Ref.~\cite{CahnActaMater04}, i.e., shifting the $\ln r$ data obtained from different simulation runs vertically in the plot due to different initial conditions yielding different $C_1$ or $C_2$. The amount of shifting is different for different sections of data (with each section referring to the data measured in a same run) as they correspond to different initial conditions. For $\theta < 30\degree$, we first fitted the lowest-angle section of $\ln r$ vs $\theta$ data into Eq.~(\ref{lnr1}) to get $C_1$ and identify the full form of master curve, and then for each subsequent data section vertically shifted all its data points by a same constant value determined by the averaged distance of $\ln r$ data in this section from the master curve (which approximates the corresponding integration constant for this data section). Similar steps were taken for $\theta > 30\degree$ with the master curve described by Eq.~(\ref{lnr2}). In those data sections showing steep descending of $\ln r(\theta)$ with large deviations from the master curve (i.e., for $\theta$ not far from $30\degree$), the vertical shifting was made instead to simply connect the subsequent sections continuously.

The resulting outcomes presented in Fig. \ref{fig:single_rotation_lnr} include three different types of initial grain setup (see Sec.~\ref{sec:setup}). A quantitative comparison of the numerical simulation data with the analytic master curves supports our qualitative analysis described above. Overall the numerical data lies along the master curves up to $\theta$ or $\pi/3-\theta$ around $15\degree$, indicating the motion of the embedded grains is dominated by coupling, with the grain rotating with increasing or decreasing angle in two coupling modes as the grain radius decreases. In the intermediate range of misorientation (i.e., $\theta$ or $\pi/3-\theta$ around $10 \degree$ to $15 \degree$), there are some deviations from the master curves, with the $\theta$ variation of $\ln r$ proceeding at a slower rate, indicating the occurrence of sliding accompanied with the coupling. At higher misorientation with $\theta$ or $\pi/3-\theta$ approaching $20\degree$, the normal motion (shrinking) and tangential translation of the grain begin to decouple, and the grain dynamics is dominated by sliding. This is indicated by the sharp deviation in the slope of the $\ln r(\theta)$ data away from the master curve predicted by Eq.~(\ref{lnr1}) or (\ref{lnr2}). In this regime $\theta$ changes only slightly while $\ln r$ decreases, corresponding to an effective rotation rate much smaller than what would be expected if the grain motion were dominated by coupling. When closer to $30\degree$ misorientation, the grain is no longer undergoing either sliding or coupling and shrinks without rotating, i.e., with $\theta$ remaining constant with time. 

Note that in our PFC simulations for single-component graphene, no dual behavior, i.e., no switching between different coupling modes during grain evolution or no dual modes at the same $\theta$, is observed. This is consistent with the previous PFC simulation of the shearing of planar grain boundary in 2D square lattice \cite{TrauttActaMater12b}, while in single-component systems so far the dual behavior has been found only in 3D MD simulations of grain boundaries \cite{CahnActaMater06,CahnPhiloMag06,ThomasNatCommun17}.

\subsection{Binary embedded grains for h-BN}
\label{sec:binary}

\begin{figure}[htbp]
    \centering
    \includegraphics[width=0.95\textwidth]{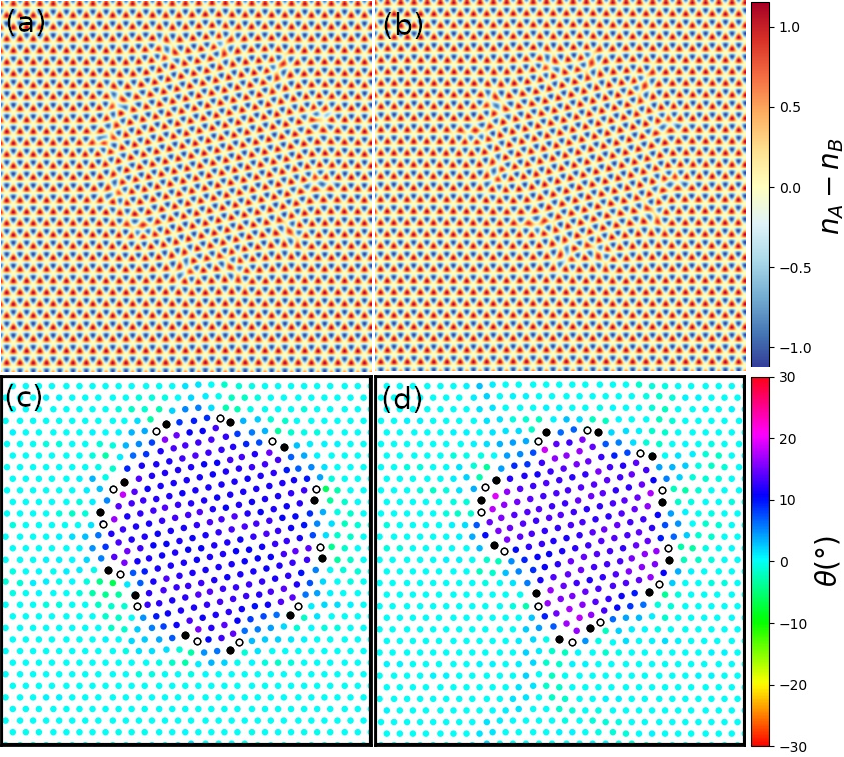}
    \caption{Simulation snapshots of a binary embedded grain during its time evolution at early [(a) and (c)] and late [(b) and (d)] times. The grain was initialized with $\theta_0 = 12 \degree$ and rotated to a larger angle. (a) and (b) show the spatial distribution of the density difference $n_A - n_B$, giving binary honeycomb structure, while (c) and (d) show the corresponding distribution of local angle $\theta$ of the same grain, with the coloring and defect marking similar to those of Fig.~\ref{fig:single_rotation}.}
    \label{fig:binary_rotation}
\end{figure}

The binary honeycomb crystalline structure with $AB$ sublattice ordering has a broken inversion symmetry due to the alternating $A$ and $B$ type atoms in the lattice, resulting in its rotational symmetry being changed from 6-fold to 3-fold, with maximum misorientation angle $\theta_{\rm max}^{\rm b} = 120 \degree$. Embedded grains in our binary PFC simulations of h-BN tended to have substantially lower mobility than their single-component counterparts. These grains would typically rotate less than $2 \degree$ or even less than $1 \degree$ before ceasing to evolve further over the timescales we investigated, with $r$ no longer decreasing, i.e., with the grains fixed in size and orientation. As described in Sec.~\ref{PFC}, here we used model parameters similar to those of single-component case, with the same temperature parameter $\epsilon_A = \epsilon_B = 0.02$. Larger values of $\epsilon$ (corresponding to lower temperature) produced even less degree of rotation and overall evolution of these binary grains. This could be attributed to the much higher Peierls barrier for dislocation motion and much stronger defect pinning effect in this binary system as compared to the single-component one, leading to more faceted interfaces and more rigid boundary motion, as seen in some simulation snapshots given in Fig.~\ref{fig:binary_rotation}.

The motion of the boundaries of binary embedded grains appeared more irregular and less rapid than the single-component case. Typically, as seen from the simulations a majority of the boundary remained stationary for an interval of time and then one or few boundary defects migrated inward. This caused the grain to rotate slightly, which then sometimes induce the motion of other defects. This intermittent and limited motion, driven by only a fraction of the total defects, would often leave the grain concave during part of its evolution, resulting in irregular grain boundary geometries as shown in Fig.~\ref{fig:binary_rotation}. 

\begin{figure}[htbp]
    \centering
    \includegraphics[width=\textwidth]{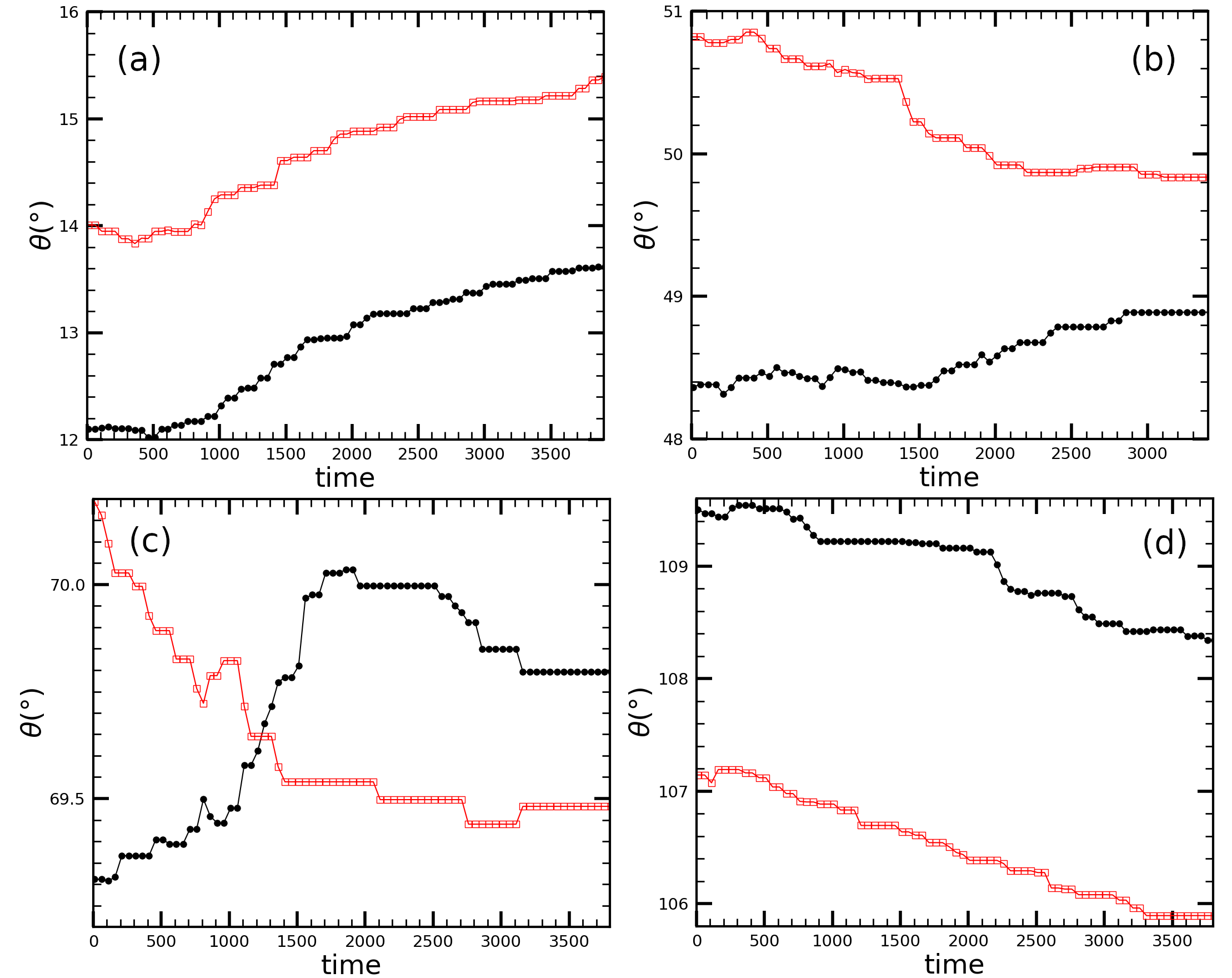}
    \caption{Time evolution of misorientation angle $\theta$ for several embedded grains in binary honeycomb PFC. (a) Grains were initialized at low angles with $0 \degree < \theta_0 < 30 \degree$, which all rotate towards increasing $\theta$ with similar rates. (b) Grains were initialized at higher angles of $30 \degree < \theta_0 < 60 \degree$, exhibiting a dual behavior. In two examples shown here, grains rotate in opposite directions despite being initialized with similar $\theta_0$. (c) Grains were initialized within the range of $60 \degree < \theta_0 < 90 \degree$, also showing the dual behavior. This dual behavior can also manifest as grains changing the direction of their rotation during the evolution, as shown in the examples here. (d) Grains were initialized within the range of $90 \degree < \theta_0 < 120 \degree$ and rotated towards decreasing $\theta$, without any dual behavior.}
    \label{fig:binary_angle_time}
\end{figure}

\begin{figure}[htbp]
    \centering
    \includegraphics[width=0.9\textwidth]{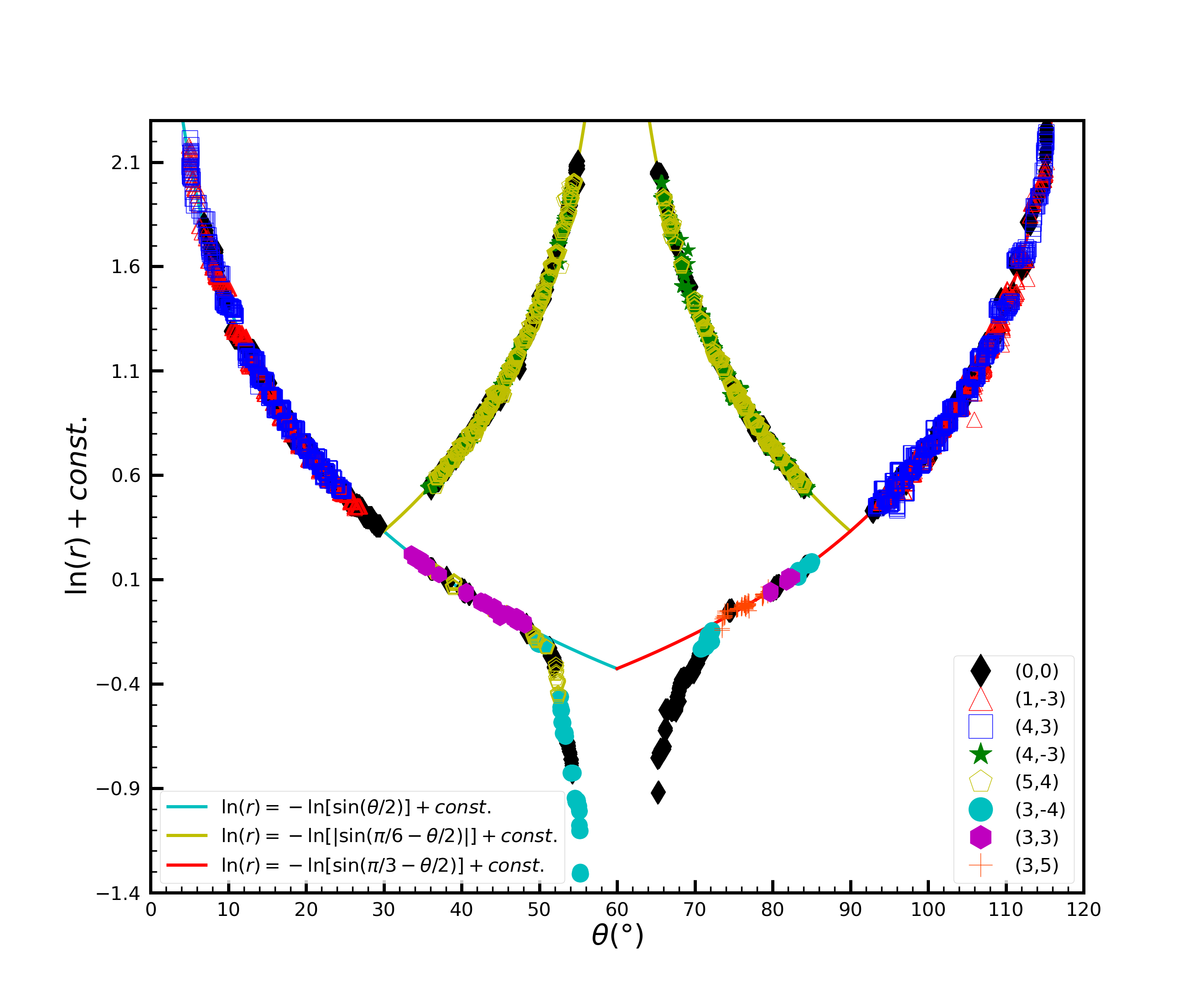}
    \caption{Plots of $\ln r$ vs $\theta$ for binary embedded grains across the full range of misorientation $0 < \theta < 120 \degree$. Both the numerical data obtained from binary PFC simulations and the analytic master curves of Eqs.~(\ref{lnr1}), (\ref{lnr2}), (\ref{lnr3}), and (\ref{lnr4}) corresponding to all three coupling branches of $\beta$ are shown. Different types of symbols represent the simulation results starting from different initial displacements of grain center with respect to the background crystal, indicated by the horizontal and vertical components of each displacement (in units of grid points). The dual behavior appears within the range of $30 \degree < \theta < 90 \degree$.}
    \label{fig:binary_rotation_all}
\end{figure}

We have conducted PFC simulations of binary embedded grains over the full misorientation range of $0\degree < \theta < 120 \degree$ and measured the time dependence of grain angle $\theta(t)$ and the corresponding effective grain radius $r(t)$. Some sample results of $\theta$ vs $t$ in four characteristic regimes of grain misorientation are shown in Fig.~\ref{fig:binary_angle_time}, and the $\ln r$ vs $\theta$ plots for all the outcomes obtained are given in Fig.~\ref{fig:binary_rotation_all}, where the numerical simulation data has been shifted vertically according to the master curves of Eqs.~(\ref{lnr1}), (\ref{lnr2}), (\ref{lnr3}), and (\ref{lnr4}) to take into account the different initial conditions of different simulation runs, in the same manner as for the single-component case described above.

At the two edges of the angle range with smallest and largest sets of $\theta$ values, i.e., $\theta < 30 \degree$ or $\theta > 90 \degree$, the binary embedded grains behaved similarly to their single-component counterparts, shrinking and rotating through a small angle towards higher or lower $\theta$, respectively, as seen in Figs.~\ref{fig:binary_angle_time}(a) and \ref{fig:binary_angle_time}(d) for some examples of time evolution of $\theta$ which increases or decreases monotonically over time, other than some small fluctuations at initial transients. They correspond to two coupling modes with positive and negative coupling factors $\beta_1$ [Eq.~(\ref{beta1})] and $\beta_3$ [Eq.~(\ref{beta3})] and the associated master curves of Eqs.~(\ref{lnr1}) and (\ref{lnr4}), as presented in the $\ln r$ vs $\theta$ results of Fig.~\ref{fig:binary_rotation_all}.
 
However, the binary grains initialized at the intermediate range of angles $30 \degree < \theta < 90 \degree$ exhibited a more complicated, dual behavior which is absent in the single-component 2D grain rotation. It showed as two forms of coupling-mode selection depending on the detailed local microstructures and energetics of elasticity and plasticity. Grains initialized with slightly different initial conditions could rotate in opposite directions, even if the initial $\theta_0$ was changed by as little as less than $1 \degree$ or the initial displacement of the grain center by only a single grid point. A related example is shown in Fig.~\ref{fig:binary_angle_time}(b). In another form of mode switching, some embedded grains initialized within this angle range would change their direction of rotation at an intermediate evolution time during a single simulation. Some examples are given in Fig.~\ref{fig:binary_angle_time}(c), where the nonmonotonic behavior of misorientation $\theta$ over time (after initial transients) is beyond the measurement errors but a result of dual behavior of grain rotation manifesting as a procedure of grain rotating, halting, and reversing of rotation direction during evolution. It is interesting to note that these PFC simulation results are consistent with those of a very recent experiment on the shrinking and rotation of h-BN grains \cite{RenACSNano20} which observed the phenomenon of dual mode switching for initial $\theta_0 \sim 35.6\degree$ and the single-coupling-mode behavior for $\theta_0 < 22\degree$. A discrepancy occurs for $38 \degree < \theta_0 < 60\degree$, where only the unidirectional rotation towards smaller $\theta$ (related to the negative branch of $\beta$ coupling) was observed experimentally, while the simulation data shown in Figs.~\ref{fig:binary_angle_time}(b) and \ref{fig:binary_rotation_all} indicates a dual behavior of two coupling modes up to $\theta \sim 50 \degree$, beyond which the grain could either rotate according to the negative coupling mode as seen in the experiment, or significantly deviates from the coupled motion with much smaller rotation. This discrepancy could be related to the limited amount of grain cases measured in the experiment and some quantitative differences in the system conditions simulated by PFC (e.g., different conditions of nanowelding method and electron irradiation used in experiments).

The systematic results of our PFC study presented in Fig.~\ref{fig:binary_rotation_all} are symmetric with respect to $\theta_{\rm max}^{\rm b}/2=60\degree$, as expected from the binary honeycomb lattice. As described above, the behavior of the binary embedded grains in the range of $30\degree < \theta < 90\degree$ can be understood not as random rotations, but as the selection or switching between dual coupling modes. Such dual behavior occurs when there are more than one accessible branches of $\beta(\theta)$ for the grain coupled motion so that both positive and negative coupling modes coexist at a given $\theta$ \cite{CahnActaMater06}, which results in the observance of both opposite directions of grain rotation. This can be shown more systematically by aligning and comparing the corresponding numerical data with the master curves of both modes in Fig.~\ref{fig:binary_rotation_all}. We separated the portions of the grain rotation data where $\theta$ monotonically changed with time, either increasing or decreasing, which correspond to separate coupling modes, and then followed the same general procedure of vertical shifting based on a master curve. Data where $\theta$ increased with time during grain rotation and shrinkage was shifted with respect to the branch of the curve with $\partial \ln r / \partial \theta < 0$ (i.e., Eq.~(\ref{lnr1}) or (\ref{lnr3})). Correspondingly, data with time-decreasing $\theta$ was shifted with respect to the other branch with $\partial \ln r / \partial \theta > 0$ (i.e., Eq.~(\ref{lnr2}) or (\ref{lnr4})), indicating the opposite direction of rotation. Dual behavior of coupling results in both branches of simulation data clearly following the respective master curves of coupled motion, as seen in Fig.~\ref{fig:binary_rotation_all}, whereas random rotation would not show any correlation between $r$ and $\theta$.

A steep departure of the simulation results of $\ln r(\theta)$ from the analytic master curves for perfect coupling could occur for $\theta$ near $\theta_{\rm max}^{\rm b}/2 = 60\degree$ (see the lower branches of numerical data in Fig.~\ref{fig:binary_rotation_all}), indicating the grains begin to experience sliding and the decoupling between normal and tangential motions (i.e., with $\beta$ descending towards zero). This is similar to the single-component case, although now it is near $60\degree$ instead of $30\degree$ due to the binary lattice symmetry. Grains initialized exactly at $60\degree$ create inversion domain boundaries, regions separating two parts of the crystal where the $A$- and $B$-type atoms are reversed. These grains tend to shrink rigidly through collective atomic displacements of defects, without rotating or sliding, as demonstrated in previous work \cite{Taha17}.

Different from the single-component results given in Fig.~\ref{fig:single_rotation_lnr}, the simulation data we obtained for binary grains closely follows the master curve right up to $\theta = 30\degree$ without steep descending of $\ln r(\theta)$, implying that the grain motions remain in the coupling mode. However, the dual behavior in its nearby range of $\theta$ caused the direction of rotation to fluctuate between opposite directions increasingly rapidly when approaching the intersection of different coupling modes at $\theta = 30\degree$ or $90\degree$, where the positive and negative branches of coupling factor $\beta(\theta)$ are nearly of the same magnitude as shown in Fig. \ref{fig:beta_branches}. It would lead to a frustrated state with almost no net rotation observed, similar to the scenario discussed in Ref.~\cite{TrauttActaMater12a}. This has been confirmed in our PFC simulations. Grains initialized with $\theta_0$ within this narrow angle range were found to experience little change in $\theta$ but with many small fluctuations over time, representing small degrees of rotations in opposite directions that frequently reversed. We were thus unable to assemble meaningful segments of rotation data within approximately $5 \degree$ of these intersection points due to this frustrated behavior (which results in the corresponding data-free regions in Fig.~\ref{fig:binary_rotation_all}).

It is noted that the simulation data presented in Fig.~\ref{fig:binary_rotation_all} appear much smoother and better aligned with the analytic master curves as compared to the single-component counterparts shown in Fig.~\ref{fig:single_rotation_lnr}, which would seem to indicate a much less degree of sliding for angles far enough from $\theta_{\rm max}^{\rm b}/2 = 60\degree$, as measured by the relative deviations from the master curves for perfect coupling. However, it is not as conclusive as it appears. Because the binary embedded grains exhibited much lower mobility and typically rotated through less than 1-2 degrees before ceasing to evolve, Fig.~\ref{fig:binary_rotation_all} is comprised of many short segments of rotation data corresponding to independent simulation runs initialized at different values of $\theta_0$. In contrast, for single-component grains much larger angle range of grain rotation during each simulation leads to longer sections of data in Fig.~\ref{fig:single_rotation_lnr}, which allows any fluctuations over time in grain size ($\ln r$) and/or misorientation $\theta$ to become apparent. The extent of such fluctuations could exceed the short range of each individual data segment in Fig.~\ref{fig:binary_rotation_all} for binary grains, and hence these fluctuations might not be effectively sampled in our binary PFC simulations. Therefore, it would be difficult to determine accurately the degree of sliding in the binary system, other than noting that the binary grain’s motion is dominated by coupling and sliding appears to play a secondary or minor role for $\theta$ far from $60\degree$. Only in the range of angles not far from $60\degree$ were the variations large enough to obviously depart from the master curves predicted by the Cahn-Taylor formulation for several sequential segments of data, resulting in the clear signal of sliding.

\section{Discussion}

The main difference between the above simulation results of single component (graphene) and binary (h-BN) 2D grain dynamics, other than the doubling of misorientation range, is the appearance of dual behavior of coupling modes for binary grain rotation that is absent in the single-component 2D system, although the lattice sites of both are of the same honeycomb structure. The angle range exhibiting this dual mode behavior is much broader than that found in the previous MD studies of 3D single-component fcc metals. Without loss of generality, in the following we consider half of the rotational symmetry period of the binary honeycomb system, $0\degree < \theta < 60\degree$, with the same analysis applied to the other half. The first part of the angle range $0 \degree < \theta < 30 \degree$ follows the same behavior as for the single-component honeycomb grains, with the coupling factor $\beta_1$ governed by Eq.~(\ref{beta1}); thus grains initialized in this range of angles rotate in a single direction towards increasing $\theta$, i.e., in the positive branch of coupling. 

To understand the dual behavior of binary grains coupled motion at misorientation angles $30\degree < \theta < 60\degree$, we need to consider factors beyond the purely geometric ones leading to Eqs.~(\ref{beta1})--(\ref{lnr4}) in the Cahn-Taylor formulation. The lattice-site structure of binary honeycomb 2D crystals like h-BN with sublattice ordering is identical to that of the single-component honeycomb (graphene), and thus for both of them the lattice sites themselves have a six-fold rotational symmetry with $\theta_{\rm max}^{\rm s}=60\degree$. This creates the identical geometric effect causing normal-translational coupled motions of the grain boundary with shear deformation, leading to unidirectional rotation of the embedded grains towards $\theta_{\rm max}^{\rm s}/2=30\degree$. In the range of $30\degree < \theta < 60\degree$ this purely geometric effect yields the negative mode of coupling (i.e., $\beta_2$ in Eq.~(\ref{beta2})), with no switching with the other positive $\beta_1$ mode found in simulations of 2D single-component grains.

Nevertheless, the binary system is not identical to their single-component counterpart. While the structure of the lattice sites is unchanged, the alternating $A$ and $B$ atomic components with sublattice ordering break the inversion symmetry of the lattice, resulting in a reduced three-fold symmetry with $\theta_{\rm max}^{\rm b} = 120\degree$. This is due to the bonding-energy difference between $A$-$B$ heteroelemental neighborings and $A$-$A$ or $B$-$B$ homoelemental ones, a key factor that has been incorporated in the PFC free energy functional Eq.~(\ref{hBN energy}). A typical example showing this inversion symmetry breaking is the occurrence of $60\degree$ grain boundaries for which the lattice planes remain continuous across the boundary without any lattice site mismatch, other than the reverse of $A$ and $B$ components. As a result of this symmetry breaking, we not only have three coupling modes that become accessible as described in Sec.~\ref{sec:multimodes} (instead of two modes in single-component systems), but also have the accessibility of the first positive branch of $\beta_1$ extended to the range of $30\degree < \theta < 60\degree$ (similarly for those beyond $60\degree$), as seen in Fig.~\ref{fig:binary_rotation_all}, leading to the competition between two coupling branches in that angle range and hence the coexistence of both two opposite directions of grain rotation. This dual behavior can be attributed to the following two factors related to the specific defect structures and energy of grain boundaries.

It has been known that the selection between different coupling modes (if they are both accessible) is mainly dependent on the detailed atomic microstructures of local dislocation defects at the grain boundary and the induced stresses to activate a specific coupling mode through its corresponding atomic mechanisms for distorting and transforming structural units at boundary \cite{CahnActaMater06}. Defects in single-component grain boundaries are purely geometric, with the change in the lattice orientation being accommodated by the corresponding changes in the individual structural units or rings. This leads to the prevalence of $5|7$ dislocation cores in single-component 2D hexagonal materials like graphene \cite{Yazyev14,Huang11,HirvonenPRB16,LiJMPS18,Zhou19}, with other types of defect motifs being rare. In 2D binary hexagonal materials like h-BN and TMDs the situation is more complex, with a much richer variety of dislocation core structures such as $4|8$, $5|7$, $4|6$, $4|10$, $6|8$, $4|4$, $8|8$, and others depending on the specific conditions of grain boundary \cite{GibbJACS13,RenNanoLett19,RenACSNano20,LinACSNano15,MendesACSNano19,Taha17}. This is related to the distinction between heteroelemental and homoelemental neighborings, and thus the formation of defect core structures containing more energetically favorable $A$-$B$ heteroelemental bonds. In addition, in the high-angle range with $30\degree < \theta < 60\degree$ the number density of defects in single-component grain boundaries decreases as $\theta$ increases due to the $60\degree$ rotational invariance, with well-separated dislocation cores at large $\theta$. However, in binary grain boundaries the number density of defects remains roughly similar within this angle range, although with the variation of specific defect structures, given that at $60 \degree$ the grain boundary exists as an inversion domain. The corresponding grain boundaries of these misorientations are composed of arrays of mostly connected dislocation cores, such that most atoms along the boundary are parts of defect (see some examples in Ref.~\cite{Taha17}). Detailed mechanisms of grain boundary motion with these connected defects should then be governed by collective local atomic displacements that are different from those of low-angle boundaries (and the single-component high-angle cases) consisting of separated or dispersed dislocations.

Consequently, during the time evolution of single-component grains most defects at the boundary would remain the same type of dislocation structure (i.e., $5|7$, as also seen in our simulations such as Fig.~\ref{fig:single_rotation}), resulting in a single atomic mechanisms for grain boundary migration as there is no structural transformation between defect types. Thus only one coupling mode can be activated for a given misorientation $\theta$, excluding the appearance of dual behavior. By contrast, in a binary system it is possible that the distortion and transformation of structural units at the grain boundary could involve different types of dynamical process given the degeneration of various types of defect ring structures. This would correspond to different atomic mechanisms of coupled grain motion, such that the boundary migration mechanism could even change during a time evolution of binary grain, as accompanied by the structural changes of some boundary defects, leading to the dual behavior of mode switching as found in our PFC simulations. The activation and change of coupling modes can be readily achieved as long as the related coupling branches are accessible, including those requiring the process of mirror symmetry breaking of local structural units, due to the diversity of the available defect configurations in the binary ordering lattice. This could also explain the much wider angle range showing dual behavior as compared to 3D single-component systems where the grain boundary migration mechanisms for mode switching are much more difficult to realize and restricted to a narrow regime near the angle of transition between two modes \cite{CahnActaMater06}.

\begin{figure}[htbp]
    \centering
    \includegraphics[width=0.95\textwidth]{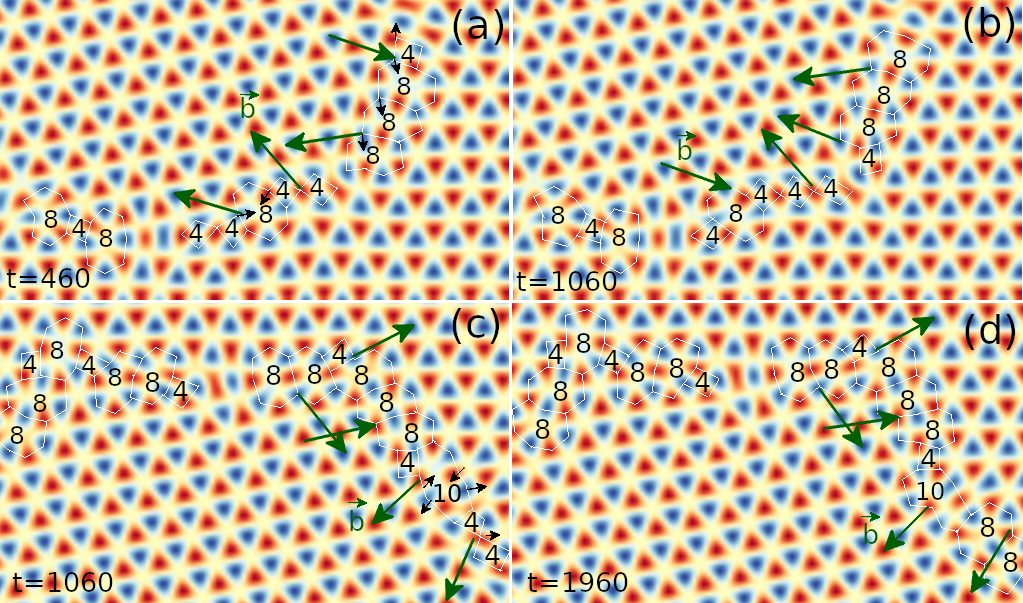}
    \caption{Simulation snapshots for two types of dislocation core transformations around a portion of the boundary of binary embedded grains initialized with $\theta_0$ close to $50 \degree$. (a) and (b) show the time evolution of a section of the lower-right quadrant of a grain that rotates from the initial $\theta_0=48.05\degree$ towards decreasing $\theta$, driven by the glide of $4|8$ or $8|8$ defect pairs. (c) and (d) show the evolution of a section of the upper-right quadrant of a grain that instead rotates from the initial $\theta_0=49.17\degree$ towards increasing $\theta$, where the grain boundary migrates through the transformation of $4|4$ defect cores together with neighboring 6-membered rings to $8|8$ dislocations, and the glide of a $10$-membered defect ring. The Burgers vectors $\vec{b}$ for defect pairs in the dislocation clusters involved in structural transformation of the grain boundary are marked with dark green arrows.}
    \label{fig:defect_figure}
\end{figure}

The above analysis is consistent with recent experimental observations of 2D h-BN monolayers \cite{RenACSNano20}, where high-resolution TEM imaging of some time-evolving h-BN samples did show different types of atomic mechanisms of grain boundary coupled motion involving dislocation glide and climb at small vs large misorientations $\theta$, associated with different dominated grain boundary defect configurations and dynamics. These included the motion of $5|7$ dislocations for low-angle misorientations, similar to that of graphene, leading to a single positive mode of shear-coupled motion as also shown above, and more complicated processes at larger $\theta$ involving $5|8|4|7$ and $5|7$ dislocations, yielding the coupled motion of another negative mode or dual behavior of mode switching \cite{RenACSNano20}. Richer types of defect transformation have been found in our PFC simulations of binary grain rotation. Some examples in the angle range showing dual behavior are given in Fig.~\ref{fig:defect_figure}, where grain boundary motions are dominated by atomic displacements of $4|8$, $4|4$, $8|8$, and $4|10$ dislocations. Rotation towards decreasing $\theta$ was associated with the glide of $4|8$ or $8|8$ dislocations (see Fig.~\ref{fig:defect_figure}(a)-(b)), whereas rotation towards increasing $\theta$ was mainly associated with chains of $4|4$ dislocations combining with nearby 6-membered rings and opening up to $4|8$ and $8|8$ dislocations and the reverse procedure, as well as the glide of $10$-membered rings (see an example shown in Fig.~\ref{fig:defect_figure}(c)-(d)). The difference between our simulation and the experimental observation of specific defect behaviors could be attributed to different setup conditions in the experiment which used a nanowelding method in the grain boundary fabrication and the electron beam irradiation to activate the grain motion (in addition to thermal activation) \cite{RenACSNano20}, as compared to the annealing process simulated here. Nevertheless, both yield qualitatively consistent results showing different types of dynamic pathways of defect evolution corresponding to different types of grain rotation and coupling modes.

Another factor facilitating dual behavior of the coupled motion would be related to the angle dependence of grain boundary energy. For single-component graphene with the full $\theta$ range of $[0\degree, 60\degree]$, the curve of the grain boundary energy $\gamma$ per unit length is known to have a shallow dip at $\theta$ slightly higher than $30\degree$ \cite{HirvonenPRB16}, which would not affect the dominance of the shear-coupled mechanism. For the binary system of h-BN with the full angle range of $[0\degree, 120\degree]$, recent calculations showed a maximum regime of $\gamma$ around $30\degree$-$40\degree$ but a steep local minima at $\theta=60\degree$ (with $\gamma$ reduced by half) \cite{Taha17}. This energetic factor could then further facilitate the accessibility of an additional $\beta_1$ or $\beta_3$ coupling mode in the range of $30\degree < \theta < 90\degree$ (the lower branches in Fig.~\ref{fig:binary_rotation_all}) with grains tending to rotate in the direction towards $60\degree$, enabling the occurrence of dual behavior of coupling modes in addition to the atomic mechanisms of defect motion described above which probably is the primary driving factor.

In the limit of $\theta$ close to $60\degree$, i.e., when $55\degree < \theta < 75 \degree$, the grain boundary structure is close to that of an inversion domain, showing as an array of connected defect rings mostly being of the $4|8$ type. Thus, the grain shrinking dynamics is similar to that of inversion domains, controlled by a chain reaction of atomic displacements of the boundary dislocation cores and the corresponding defect shape transformation, without bulk motion of the grain \cite{Taha17}. This then results in the absence of any grain rotation, sliding, or coupled motion, with $\theta$ kept unchanged during grain shrinkage, corresponding to the mid region of Fig.~\ref{fig:binary_rotation_all}. In the other limit of $\theta$ very close to $30\degree$ or $90\degree$, the absence of net grain rotation is due to the corresponding frustrated state explained above in Sec.~\ref{sec:binary}. This is caused by the similar degree of coupling factor $\beta$ for two positive and negative modes, such that both atomic mechanisms of defect motion leading to opposite rotation directions can occur almost equally likely at a given moment during a time evolution of grain.

\section{Conclusions}

We have conducted a systematic study on the dynamics of single- and two-component grains and the coupled grain boundary motion in 2D crystals with honeycomb lattice symmetry, through the PFC modeling of graphene and h-BN type 2D hexagonal materials. By tracking the motion of time-evolving embedded grains misoriented with respect to the surrounding crystalline matrix, we have analyzed the angle dependence of grain boundary dynamics and its implications for grain growth across the full range of misorientations. Our results indicate that over the majority of misorientation angles $\theta$ the behaviors of both single-component and binary 2D grains are governed by the coupling between normal and tangential motions of the grain boundary, well following the Cahn-Taylor formulation in terms of the $\theta$ dependence of the coupling factor $\beta(\theta)$ and the grain radius $r(\theta)$ as given in Eqs.~(\ref{beta1})--(\ref{lnr4}). This is seen from both the simulation outcomes of grain rotation direction and the matching of numerical data to the analytic master curves for perfect coupling. The coupling is weakened when $\theta$ approaches the mid point of the maximum allowed misorientation angle (i.e., $30\degree$ for single-component and $60\degree$ for binary honeycomb lattice), as accompanied by the occurrence or dominance of grain boundary sliding. When in the closer vicinity of the mid angle, no angle change is observed during grain shrinking, indicating the absence of both coupling and sliding.

One of our key findings is the occurrence of dual behavior of grain coupled motion within the misorientation range of $30\degree < \theta < 90\degree$ in the binary 2D hexagonal system, which is missing in the 2D single-component counterpart. It manifests as grains being able to rotate in both of the opposite directions towards higher and lower angles corresponding to two different coupling modes. Our systematic study predicts a broad range of misorientation angles exhibiting the dual behavior of mode selection or switching, much broader than that observed previously in 3D simulations of single-component systems \cite{CahnActaMater06,CahnPhiloMag06,ThomasNatCommun17} and experiments of 2D h-BN monolayers \cite{RenACSNano20}, as a result of a much more diverse family of defect core structures and dynamics available in the binary 2D materials. This dual behavior in 2D binary systems is originated from the coexistence of both positive and negative branches of coupling in this angle range, of which the additional mode associated with the grain rotation to a direction towards $\theta=60\degree$ is made accessible through different dominated types of boundary defect configurations and defect evolution pathway, as further facilitated by the sharp local minimum of grain boundary energy at $60 \degree$ misorientation. The activation, selection, and switching of different coupling modes are then determined by the specific dislocation core microstructures and arrangements at the embedded grain boundary and their structural transformations during the grain evolution. The origin of the availability of a rich variety of dislocation structures enabling the 2D dual behavior is the breaking of inversion symmetry in the binary honeycomb lattice with $AB$ sublattice ordering, given the energetic difference between $A$-$B$ heteroelemental and $A$-$A$ or $B$-$B$ homoelemental bondings, a pivotal factor that is absent in single-component systems. 

Similar material systems, where the crystalline structure has a symmetry breaking because of alternating lattice sites occupied by different atomic species, should be expected to exhibit similar dual behavior of coupling. This would be important for the understanding and control of grain growth mechanisms and hence the production of large-scale single crystals, not only for 2D materials like h-BN that is modeled here but also for other binary materials such as TMDs with similar structure of sublattice ordering and defect configurations as well as other 2D or 3D multi-component material systems. These materials would be governed by more complex dynamics of grain evolution when subjected to annealing as compared to single-component ones, as a result of the competing mechanisms involved particularly those caused by dual behavior of coupled grain boundary motion and collective atomic displacements for various types of defect transformations. All these emphasize the need for further studies to explore more detailed atomic mechanisms of grain rotation and dynamics in multi-component systems under a wider range of growth and processing conditions.

\section*{Acknowledgements}

This work was supported by the National Science Foundation under Grant No. DMR-2006446. We thank Talbot Knighton for the help on the coding implementation of angle calculations based on Delaunay triangulation.

\bibliographystyle{elsarticle-num}
\bibliography{hBN_grainrotation_refs}

\end{document}